\documentclass{epl}
\usepackage{graphicx}
\usepackage{amsmath}
\usepackage{amsfonts}
\usepackage{amssymb}

\title{Constructing finite dimensional codes  \\
with optical continuous variables}
\author{S. Pirandola \and S. Mancini \and D. Vitali \and P. Tombesi}
\institute{INFM, Dipartimento di Fisica, Universit\`a di
Camerino, I-62032 Camerino, Italy}


\begin{document}

\maketitle

\begin{abstract}
We show how a qubit can be fault-tolerantly encoded in the infinite-dimensional Hilbert space
of an optical mode. The scheme is efficient and realizable with present technologies.
In fact, it involves two travelling optical modes coupled by a cross-Kerr interaction, initially
prepared in coherent states, one of which is much more intense than the other.
At the exit of the Kerr medium, the weak mode is
subject to a homodyne measurement and a quantum codeword is conditionally generated in the
quantum fluctuations of the intense mode.
\end{abstract}

\section{Introduction}

Quantum Information has recently entered
the domain of continuous variable (CV) systems \cite{vaidman,lloyd1} and
Quantum Error Correction theory in particular has been extended to
the CV framework \cite{Lloyd2,preskill}.
In such a framework, the main damage induced by decoherence is typically a
small diffusion in the two canonical coordinates characterizing the CV system, which
may represent the position and momentum of a particle or two orthogonal field quadratures
of an optical mode.
Gottesman \emph{et al.} have shown that these effects are best fought
by encoding quantum information within appropriate \emph{shift-resistant} quantum codes of the CV system \cite{preskill}.
The relevant aspect of the proposal of Ref.~\cite{preskill} is that, using qubits encoded in such a way, a
universal set of fault-tolerant quantum gates can be implemented using only ``cheap'' standard devices which, in the case
of a CV optical mode, correspond to beam splitters, phase shifters, squeezers, homodyne and photodetectors.
In such a proposal, all the difficulties are confined to the preparation of the encoded states which, in the ideal limit, are superpositions of an infinite number
of infinitely squeezed states and require the use of particular nonlinear interactions for their generation.
The preparation of these shift-resistant quantum codes represents the
main drawback of the Gottesman \emph{et al.} proposal for universal fault-tolerant quantum computation in a CV system and therefore
it is crucial to find an experimentally viable way to generate the quantum codewords employing optical CV modes.
A first attempt has been based on a conditional scheme in which the CV system is coupled to a single qubit
and which could be implemented on a single trapped ion \cite{travaglione}.
Here, elaborating on a suggestion in \cite{preskill} of using a
nonlinear Hamiltonian of ponderomotive type, we present an all-optical scheme involving
two modes interacting in a crossed-Kerr medium.

\section{Ideal and approximate codewords}

\begin{figure}[ptbh]
\begin{center}
\includegraphics[width=0.7\textwidth]{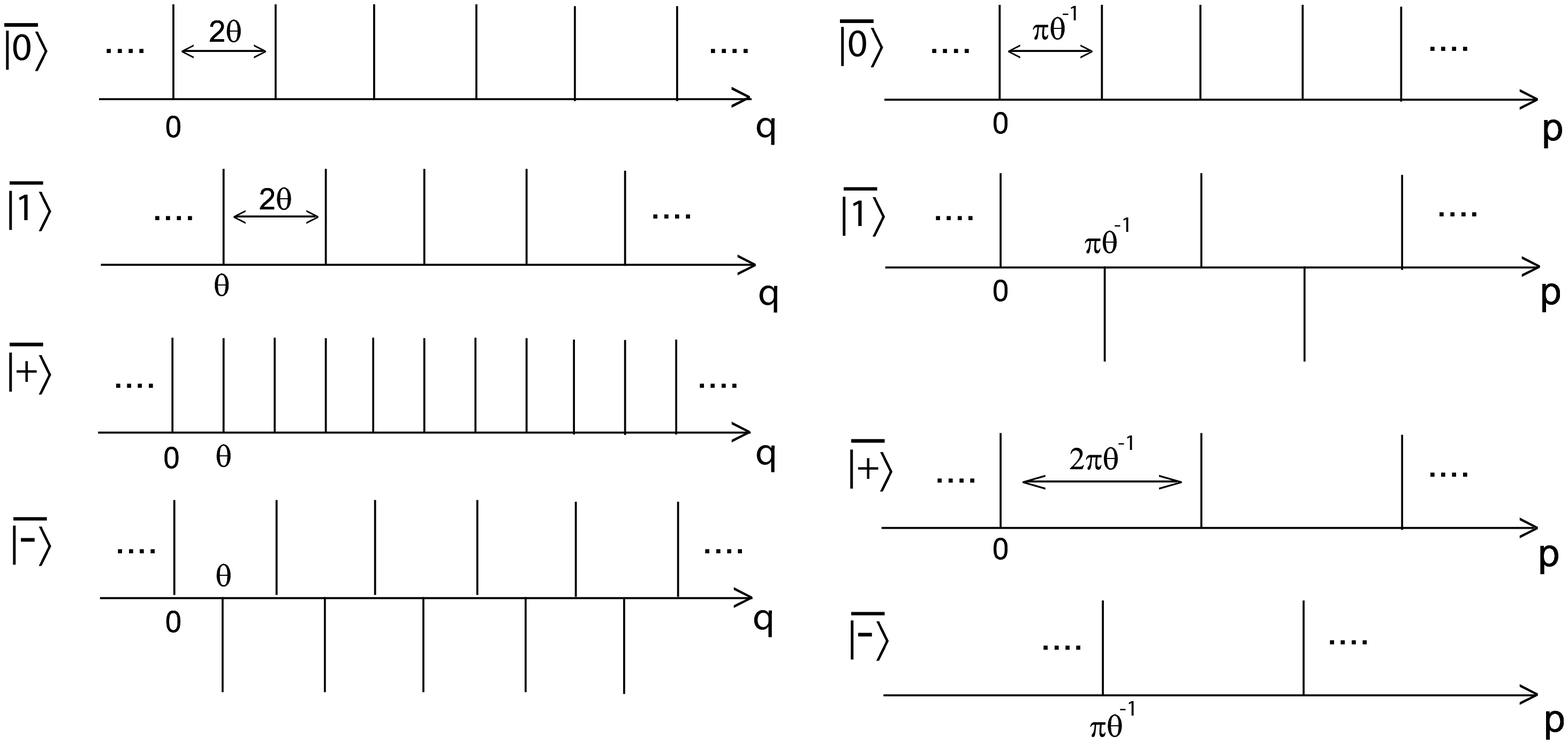}
\end{center}
\par
\vspace{-0.5cm}\caption{Schematic representation of the position and momentum wave functions of the four ideal codewords
$\overline{|  0\rangle}$, $\overline{|  1 \rangle}$, $\overline{|  + \rangle}$ and $\overline{|  -\rangle }$.} \label{ideal}
\end{figure}

Ref.~\cite{preskill} shows that a single qubit, whose generic state can be written as $c_{0}\left|  0\right\rangle +c_{1}\left|
1\right\rangle $, can be encoded in an oscillator with canonical coordinates $\hat{q},\hat{p}$ (with $[\hat{q},\hat{p}]=i$) so that the
corresponding logical state $c_{0} \overline{\left|  0\right\rangle
}+c_{1}\overline{\left|  1\right\rangle }$ is easily correctable with
respect to small diffusion errors in both position and momentum.
The codewords $\overline{\left|  0\right\rangle }$ and $\overline{\left|  1\right\rangle
}$ are the simultaneous eigenstates with eigenvalue $+1$ of the two
displacement operators $\hat{T}_{q}(2\theta)= \exp\left\{-2i\hat{p}\theta\right\}$ and $\hat{T}_{p}(2\pi\theta^{-1})=\exp\left\{2i\pi\hat{q}/\theta\right\}$
(the so called \emph{stabilizer generators} of the code \cite{stabilizer}) with
arbitrary $\theta\in\mathbb{R}$.
The two codewords are a coherent superposition of infinitely squeezed states
both in position and momentum, that is,
\begin{eqnarray}
&& \overline{\left|  0\right\rangle
}=\sum_{s=-\infty}^{+\infty}\left|  q=2\theta s\right\rangle
=\sum_{s=-\infty}^{+\infty}\left|  p=\pi\theta^{-1} s\right\rangle
\label{ideal_0} \\
&& \overline{\left|  1\right\rangle
}=\hat{T}_{q}(\theta)\overline{\left| 0\right\rangle
}=\sum_{s=-\infty}^{+\infty}\left|  q=2\theta s+\theta
\right\rangle =\sum_{s=-\infty}^{+\infty}(-1)^{s}\left|
p=\pi\theta ^{-1}s\right\rangle \label{ideal_1}.
\end{eqnarray}
They have a comb-like structure with spacing $2\theta$ in the $q$-space and $\pi/\theta$ in the $p$-space and
are displaced with respect to each other by $\theta$ along the $q$-axis (this is also responsible for the alternating
signs in the p-representation, see Fig.~\ref{ideal}).

In a CV system, errors and decoherence manifest themselves through random shifts of the two quadratures $\hat{q}$ and $\hat{p}$. The present coding
is designed for the case when these shifts are never too large, i.e., the error is correctable only if the shift is below a certain threshold.
In fact, the recovery process is realized by measuring the stabilizer generators $\hat
{T}_{q}(2\theta)$, which is equivalent to measure $\hat{p}$ ($\operatorname{mod}\pi\theta
^{-1}$) and therefore detects momentum shifts, and $\hat{T}_{p}(2\pi\theta^{-1})$,
which is equivalent to measure $\hat{q}$ ($\operatorname{mod}\theta$) and detects position shifts.
Then the error is corrected by displacing the state to the nearest peak, that is, to the nearest multiple of $\pi\theta^{-1}$
along the $\hat{p}$-axis and to the nearest multiple of $\theta$ along the $\hat{q}$-axis. Therefore the recovery is successful
only when the momentum shift $ \Delta p$ satisfies $\left|  \Delta p\right|  <\pi\theta^{-1}/2$ and the position shift
$\Delta q$ satisfies $\left|  \Delta q\right|  <\theta/2$, because otherwise the displacement would recover the wrong
codeword, corrupting therefore the encoded information. An alternative choice for the codewords is the basis formed by the two
superposition states $\overline{|\pm \rangle }\equiv(\overline{|  0 \rangle }\pm\overline{|  1\rangle })/\sqrt{2}$, which are used in \cite{preskill}
for the error recovery in the $p$-space because they have the same displaced comb-like structure of $\overline{|  0 \rangle}$ and $\overline{|  1\rangle}$
in the $p$-space rather than in the $q$-space (see Fig.~1). Practically, the two sets of codewords $\{\overline{|0\rangle},
\overline{|1\rangle}\}$, $\{\overline{|+\rangle},\overline{|-\rangle}\}$
are needed to measure the stabilizers $\hat
{T}_{q}(2\theta)$ and $\hat{T}_{p}(2\pi\theta^{-1})$ respectively \cite{preskill}.

In real experiments one can only create finite squeezing and, therefore,
only approximate codewords, $\widetilde{\left|  1\right\rangle }\sim\overline
{\left|  1\right\rangle }$, $\widetilde{\left|  0\right\rangle }\equiv
\hat{T}_{q}(\theta)\widetilde{\left|  1\right\rangle }$ and $\widetilde{\left|  \pm\right\rangle
}\equiv(\widetilde {\left|  0\right\rangle }\pm\widetilde{\left|
1\right\rangle })/\mathcal{N} _{\pm}$ with
$\mathcal{N}_{\pm}=\{2(1\pm\operatorname{Re}[\widetilde
{\left\langle 0\right.  }\widetilde{\left|  1\right\rangle
}])\}^{1/2}$. In this practical situation, even if we had only small position and momentum shifts satisfying the above conditions
for a correct recovery, one would still have errors due to the use of approximate codewords possessing a nonzero overlap $\widetilde{\left\langle 0\right. }\widetilde{\left| 1\right\rangle}$
and
$\widetilde{\left\langle +\right. }\widetilde{\left| -\right\rangle}$. These \emph{intrinsic errors} manifest themselves once the measurement
of a stabilizer is performed.
The quality of an experimental implementation of the error correction scheme of \cite{preskill} is therefore established
by an \emph{intrinsic error probability}, which can be considered as the largest error probability among those associated with the four approximate codewords
$\widetilde{| 0 \rangle}$, $\widetilde{| 1 \rangle}$, $\widetilde{| \pm \rangle }$.

\section{The all-optical implementation scheme}

\begin{figure}[ptbh]
\begin{center}
\includegraphics[width=0.5\textwidth]{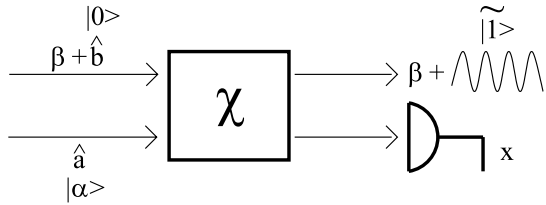}
\end{center}
\par
\vspace{-0.5cm}\caption{Schematic description of the all-optical CV encoding scheme.} \label{setup}
\end{figure}

Our proposal for encoding a qubit in a CV optical mode is schematically described in Fig.~\ref{setup}.
We consider two radiation modes $\hat{a}$ and $\hat{b}$ ($[\hat{a},\hat{a}^{\dagger}]=[\hat{b},\hat{b}^{\dagger}]=1$) impinging
on a nonlinear medium where they interact through the cross-Kerr effect, described by the interaction Hamiltonian \cite{imoto}
\begin{equation}
\hat{H}=\hbar\chi\hat{a}^{\dagger}\hat{a}\hat{b}^{\dagger}\hat
{b},\label{Ham1}
\end{equation}
where $\chi$\ is the cross-Kerr susceptibility of the medium.
The two modes are both prepared in coherent states, $|\alpha \rangle $ and $|\beta \rangle $, with $\beta \gg \alpha $ (we take $\alpha$ and $\beta$
real from now on).
We are interested in the fluctuations of the $\hat{b}$ mode around the classical intense coherent state $|\beta \rangle $, and therefore,
redefining the $b$ mode through the displacement $\hat{b}\rightarrow\hat{b}+\beta$ and neglecting
higher order contributions, the interaction Hamiltonian of Eq.~(\ref{Ham1}) can be rewritten as
\begin{equation}
\hat{H}=\hbar\chi\beta\hat{a}^{\dagger}\hat{a}(\hat{b}+\hat{b}^{\dagger
}) = -\hbar k\hat{a}^{\dagger}\hat{a}(\hat{b}+\hat{b}^{\dagger
}). \label{Ham2}
\end{equation}

If the losses within the Kerr medium are negligible the system dynamics is unitary and one has that the two-mode state after an interaction time
$t$ within the Kerr medium is given by
\begin{equation}
\left|  \phi(t)\right\rangle =\pi^{-1}\int d^{2}\gamma
e^{-(\alpha^{2}+\left| \gamma\right|
^{2})/2}\sum_{n=0}^{\infty}\frac{\alpha^{n}}{\sqrt{n!}}
\exp\{nkt[i\gamma^{\ast}-nkt/2]\}\left|  n\right\rangle _{a}\left|
\gamma\right\rangle _{b}\label{Evol2}
\end{equation}
where $\left|  n\right\rangle _{a}$ and $\left|  \gamma\right\rangle _{b}$ respectively represent
a number state for $\hat{a}$ and a coherent state for $\hat{b}$.

Just at the exit of the Kerr medium, mode $\hat{a}$ is subject to a homodyne measurement of the field quadrature
$\hat{x}=(\hat{a}+\hat{a}^{\dagger})/\sqrt{2}$ with measurement result $x$.
As a consequence, the two modes are disentangled, with mode $\hat{a}$ in the corresponding eigenstate $\left|
x\right\rangle _{a}$ of $\hat{x}$ and mode $\hat{b}$
in the state conditioned to the measurement result $x$, which, introducing the quadratures
$\hat{q}=(\hat{b}+\hat{b}^{\dagger})/\sqrt{2}$ and
$\hat{p}=(\hat{b}-\hat{b}^{\dagger})/i\sqrt{2}$, can be
described by the spatial and momentum wave functions
\begin{eqnarray}
\varphi(q;\tau,\alpha,x)&=& \pi^{-1/2}\mathcal{N}\sum_{n=0}^{\infty}\mu_{n}
(\alpha,x)\exp[-\tfrac{1}{2}q^{2}+i\pi n \tau q] \label{waveq} \\
\psi(p;\tau,\alpha,x)&=& \pi^{-1/2}\mathcal{N}\sum_{n=0}^{\infty}\mu_{n}
(\alpha,x)\exp[-\tfrac{1}{2}(p-\pi n\tau)^{2}], \label{wavep}
\end{eqnarray}
where
$
\mathcal{N}=\pi^{1/4}\{\sum_{n,n^{\prime}=0}^{\infty
}\mu_{n}\mu_{n^{\prime}}e^{-\frac{\pi^{2}}{4}(n-n^{\prime})^{2}\tau^{2}}\}^{-1/2}
$, $\mu_{n}(\alpha,x)\equiv\rho_{n}(\alpha,x)e^{-(\alpha^{2}+x^{2})/2}$,
$\rho_{n}(\alpha,x)\equiv \alpha^{n}H_{n}(x)/2^{n/2}n!$,
with $H_{n}(x)$ the Hermite polynomials, and we have introduced a dimensionless, scaled, interaction time
$\tau = \sqrt{2} k t/\pi$.
Our proposal consists in taking this conditional state as the approximate codeword
$\widetilde{|  1 \rangle }$, i.e., $\varphi(q;\tau,\alpha,x)=\langle
q  \widetilde{|  1 \rangle }$ and $\psi(p;\tau,\alpha
,x)=\langle p  \widetilde{|  1 \rangle }$, with the following correspondence of parameters $\theta \leftrightarrow 1/2\tau$.
The other approximate codeword state $\widetilde{|  0 \rangle }$ is then obtained by applying a displacement, i.e.,
$\widetilde{ | 0 \rangle }\equiv
\hat{T}_{q}(1/2\tau)\widetilde{ |1 \rangle }$, while $\widetilde{|  \pm \rangle }$ are obtained using the fault-tolerant schemes
discussed in \cite{preskill}.

\section{Evaluating the performance of the encoding scheme \label{nullo}}

From the wave functions of the encoded states shown in Fig.~1 it is easy to define and calculate the intrinsic error probabilities.
In fact, an intrinsic error occurs when: $i)$ given $\widetilde{\left|  1\right\rangle }$, the measurement of $\hat
{q}$ (mod $\theta$) gives a result inside one of the error
regions $R_{s}^{1}\equiv [ (2s-1/2)\theta,(2s + 1/2)\theta]$ for integer $s$; $ii)$
given $\widetilde{\left|  0\right\rangle }$, the measurement of $\hat{q}$
(mod $\theta$) gives a result inside one of the error
regions $R_{s}^{0}\equiv [ (2s+1/2)\theta,(2s + 3/2)\theta]$;
$iii)$ given $\widetilde{\left| - \right\rangle }$, the measurement of $\hat
{p}$ (mod $\pi/\theta $) gives a result inside one of the
error regions $R_{s}^{-}
\equiv[ (2s-1/2)(\pi/\theta),(2s + 1/2)(\pi/\theta)]$;
$iv)$ given $\widetilde{\left| + \right\rangle }$, the measurement of $\hat
{p}$ (mod $\pi/\theta $) gives a result inside one of the
error regions $R_{s}^{+}
\equiv[ (2s+1/2)(\pi/\theta),(2s + 3/2)(\pi/\theta)]$.
Therefore, the corresponding intrinsic error probabilities are given by
$\Pi_{q}(\tau,\alpha,x)\equiv\sum_{s=-\infty}^{+\infty}\int_{R_{s}^{1} }
|\left\langle
q\right. \widetilde{\left| 1\right\rangle }|^{2}dq$, for both cases $i)$ and $ii)$ and by
$\Pi_{\pm}\equiv\sum_{s=-\infty}^{+\infty}
\int_{R_{s}^{\pm}}|\left\langle p\right. \widetilde{\left|
\pm\right\rangle }|^{2}dp$ for cases $iii)$ and $iv)$. As discussed above, we can take as the overall
intrinsic error probability of the proposed approximate encoding the largest error probability,
that is, $\Pi_{\max}(\tau,\alpha,x)\equiv\max[\Pi_{q}(\tau,\alpha
,x),\Pi_{-}(\tau,\alpha,x),\Pi_{+}(\tau,\alpha,x)]$.
It can be seen that this quantity is a decreasing function of the scaled interaction time $\tau$, at least for sufficiently large values of
$\tau$. It is therefore convenient to choose a large $\tau$ (which means using a large enough value of
the coherent amplitude $\beta$) and simply consider the asymptotic value $\Pi_{\max}^{\infty}(\alpha,x)\equiv lim_{\tau\to \infty} \Pi_{\max}(\tau,\alpha,x)$.
The threshold value $\tau_{th}$ of the scaled interaction time beyond which the asymptotic values of the error probabilities
are essentially reached can be easily evaluated; $\tau_{th}$ depends upon $\alpha$ and for $0.5 < \alpha < 2.5$ it is $\tau_{th} \simeq 2$ and therefore
easily achievable with a moderate value of $\beta$.

\begin{figure}[ptbh]
\begin{center}
\includegraphics[width=0.9\textwidth]{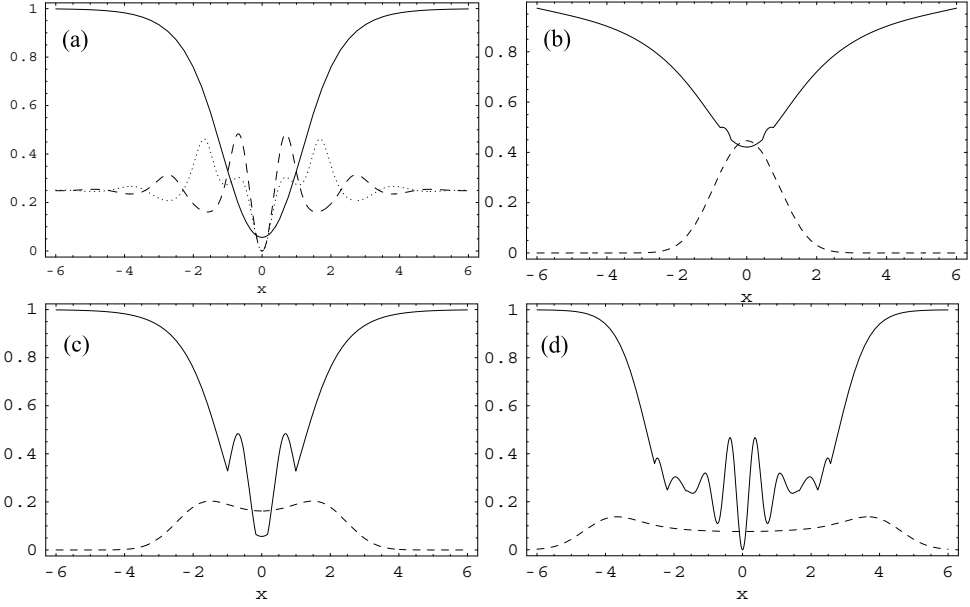}
\end{center}
\par
\vspace{-2cm}\caption{(a) Asymptotic error probabilities
$\Pi_{q}^{\infty}(\alpha,x)$ (solid line), $\Pi_{-}^{\infty}(\alpha,x)$ (dashed
line),$\ \Pi_{+}^{\infty}(\alpha,x)$ (dotted line) versus the homodyne measurement result $x$ and
for $\alpha=1.5$. The error probability $\Pi_{\max}^{\infty}(\alpha,x)$ is
given by the largest value between these three functions and its behavior for three different values of $\alpha$
is shown (solid line) in (b) ($\alpha =0.5 $), (c) ($\alpha =1.5 $) and (d) ($\alpha =3 $).
The dashed line in (b), (c) and (d) corresponds to the asymptotic probability density, $P^{\infty}(x,\alpha)$, of getting the result
$x$.} \label{probs15}
\end{figure}

The asymptotic error probabilities $\Pi_{q}^{\infty}(\alpha,x)$ and $\Pi_{\pm}^{\infty}(\alpha,x)$ can be evaluated analytically and are given by
\begin{eqnarray}
\Pi_{q}^{\infty}(\alpha,x)&=& \frac{1}{2}+\frac{2}{\pi}\frac{\sum_{n,k=0}
^{\infty}(-1)^{k} (2k+1)^{-1}\rho_{n}\rho_{n+2(2k+1)}}{\sum_{n=0}^{\infty}
\rho_{n}^{2}} \label{PAIq_x} \\
\Pi_{\pm}^{\infty}(\alpha,x)&=& \frac
{\sum_{k=0}^{\infty}\rho_{2k+1}^{2}}{2\sum_{k=0}^{\infty}(\rho_{2k+1}
^{2}+2\rho_{4k+(1\mp1)}^{2})} \label{PAIp_x}
\end{eqnarray}
and their behavior as a function of the homodyne measurement result $x$ is shown in Fig.~\ref{probs15}a.

\begin{figure}[ptbh]
\begin{center}
\includegraphics[width=0.8\textwidth]{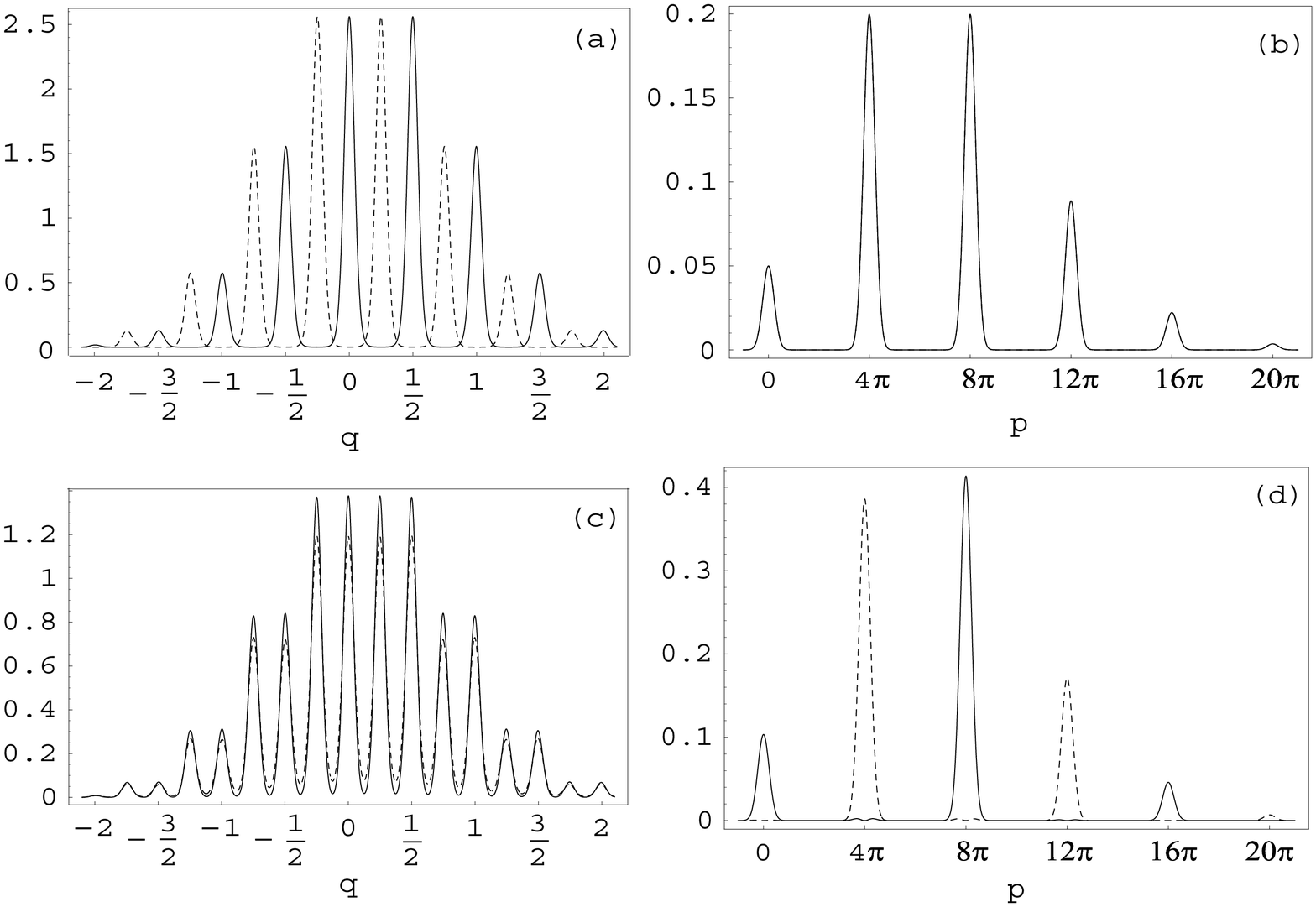}
\end{center}
\par
\vspace{-0.5cm}\caption{Probability densities of the approximate codewords
$\widetilde{\left| 0\right\rangle }$, $\widetilde{\left|
1\right\rangle }$, $\widetilde{\left| \pm\right\rangle
}$ for $\alpha=\tau=2$ ($\theta=1/4$). In (a) the
probability densities $|\left\langle q\right.
\widetilde{\left|  0\right\rangle }|^{2}$ (solid line) and
$|\left\langle q\right.  \widetilde{\left|  1\right\rangle }|^{2}$
(dashed line) are shifted by $\theta=1/4$ as expected, while in (b)
the corresponding probability densities $|\left\langle p\right.
\widetilde{\left| 0\right\rangle }|^{2}$ and $|\left\langle
p\right.  \widetilde{\left| 1\right\rangle }|^{2}$ (solid line)
coincide as expected, with spikes at $\pi\theta^{-1}s=4\pi s$.
In (c) the probability densities
$|\left\langle q\right.  \widetilde{\left|  +\right\rangle }|^{2}$
(solid line) and $|\left\langle q\right.  \widetilde{\left|
-\right\rangle }|^{2}$ (dashed line) differ only slightly, with
spikes at $\theta s=s/4$ while in (d) their momentum probability
densities $|\left\langle p\right.  \widetilde{\left|
+\right\rangle }|^{2}$ (solid line) and $|\left\langle p\right.
\widetilde{\left|  -\right\rangle }|^{2}$ (dashed line) are
approximately shifted by $\pi\theta^{-1}=4\pi$.} \label{fonda}
\end{figure}

It is evident from Fig.~\ref{probs15} that the lowest intrinsic error is achieved for measurement results $x$ as close as possible to $x=0$,
which is the measurement result for which $\Pi_{\max}^{\infty}(\alpha,x)$ is minimum.
The probability densities in $q$ and
$p$ for the four approximate codewords $\widetilde{\left|
0\right\rangle }$, $\widetilde{\left|  1\right\rangle }$, and $\widetilde{\left| \pm \right\rangle}$
corresponding to the particular measurement result $x=0$ are shown in Fig.~\ref{fonda}, where the
parameter values $\alpha=\tau=2$ have been chosen. A first qualitative comparison with the ideal codewords of Fig.~1, shows that
the present scheme is able to generate well approximated codewords.

\begin{figure}[ptbh]
\begin{center}
\includegraphics[width=0.9\textwidth]{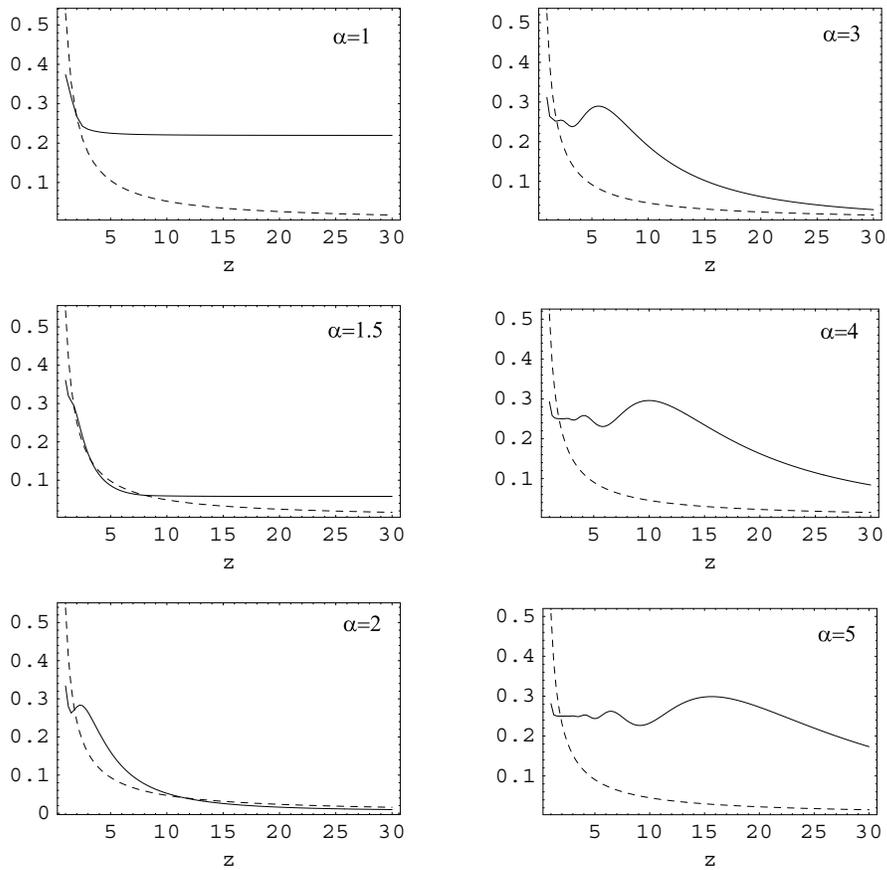}
\end{center}
\par
\vspace{-1cm}\caption{Success probability $P(z,\alpha)$ (dashed
line) and mean intrinsic error probability $\Pi(z,\alpha)$ (solid
line) versus $z$ for $\alpha=1,1.5,2,3,4,5$.}
\label{duobis}
\end{figure}

It is also evident from Fig.~\ref{probs15} that for values of $x$ of the order or larger than $\alpha$
the error probability becomes unacceptably large. Therefore the coding scheme proceeds by fixing an acceptable interval
centered around $x=0$ for the homodyne result, say $-\alpha/z < x < \alpha/z$, with $z \geq
1 $, and accepting the resulting approximate
codeword $\widetilde{\left|  1\right\rangle }$
only if $x$ lies within this interval. In such a case one has two relevant quantities:
$i)$ the \emph{success probability} of the coding scheme
$P(z,\alpha)=\int_{-\alpha/z }^{+\alpha/z}P^{\infty}(x,\alpha)dx$, where
$P^{\infty}(\alpha,x)= \pi^{-1/2}\sum_{n=0}^{\infty}\mu_n^2$ is the probability density of getting the result
$x$ in the homodyne measurement (referred to the asymptotic limit of large $\tau$);
$ii)$ the corresponding \emph{mean intrinsic error probability}
$\Pi(z,\alpha)=\int_{-\alpha/z}^{+\alpha/z}P^{\infty}
(x,\alpha)P(z,\alpha)^{-1}\Pi_{\max}^{\infty}(\alpha,x)dx$. The success probability $P(z,\alpha)$ gives the fraction of times
a satisfactory approximate codeword is generated. The mean intrinsic
error probability $\Pi(z,\alpha)$ is the most relevant quantity since it measures the effective quality of the generated codewords.
The behavior of these two probabilities is plotted as a function of $z$ for different values of $\alpha$ in Fig.~\ref{duobis}.
It is easy to see that for $z=0$, which means accepting all measurement results, $\Pi(0,\alpha)\geq 1/2$, i.e.
the encoding is too bad and cannot be used for quantum information processing.
It is also evident that the best results are achieved for vary large $z$, i.e., when only values very close to $x=0$ are accepted,
where however the success probability becomes lower.
In general the value of the coherent amplitude $\alpha$ has not to be either too large or too small. In fact, if it is too small, the intrinsic error
probability is always too large (see Fig.~\ref{duobis} for $\alpha =1$), while if it is too large, $\Pi(z,\alpha)$ stays around the value
$\Pi(z,\alpha)=1/4$ for a large interval of $z$ (see Fig.~\ref{duobis} for $\alpha = 3,4,5$) due to the oscillating behavior of $\Pi_{\max}^{\infty}(\alpha,x)$
(see Figs.~\ref{probs15}(c) and \ref{probs15}(d)). This fact implies that one has to use very large values of $z$ to achieve an acceptable intrinsic error,
corresponding however to a very low success probability. We see from Fig.~\ref{duobis} that one has a good compromise between success probability
and intrinsic error probability for intermediate $\alpha$, such as, for example, $1.5 \leq \alpha \leq 2 $. As an example, when $\alpha=2$ and $z \simeq 27$ one has
a mean intrinsic error probability $\Pi(27,2)\sim 1.0\%$ and a success probability $P(27,2)\sim 1.7\%$.

\section{Conclusions}

We have presented an all-optical conditional scheme able to generate the quantum codewords
of the CV encoding scheme proposed in Ref.~\cite{preskill} for fault tolerant quantum computation.
The generation of the encoded states is the most difficult part of that proposal and necessarily involves
a nonlinear interaction. The present scheme involves two travelling beams initially prepared in coherent states
interacting in a nonlinear cross-Kerr medium and is feasible with present technologies.

\end{document}